# Excitons, trions and Rydberg states in monolayer MoS$_2$ revealed by low-temperature photocurrent spectroscopy

*Daniel Vaquero[1], Vito Clericò[1], Juan Salvador-Sánchez[1], Adrián Martín-Ramos[1], Elena Díaz[2], Francisco Domínguez-Adame[2], Yahya M. Meziani[1], Enrique Diez[1] and Jorge Quereda[1]\**

[1] Nanotechnology Group, USAL–Nanolab, Universidad de Salamanca, E-37008 Salamanca, Spain
[2] GISC, Departamento de Física de Materiales, Universidad Complutense, E-28040 Madrid, Spain

\* e-mail: j.quereda@usal.es

**Abstract:** Exciton physics in two-dimensional semiconductors are typically studied by photoluminescence spectroscopy. However, this technique does not allow for direct observation of non-radiating excitonic transitions. Here, we use low-temperature photocurrent spectroscopy as an alternative technique to investigate excitonic transitions in a high-quality monolayer MoS$_2$ phototransistor. The resulting spectra presents excitonic peaks with linewidths as low as 8 meV. We identify spectral features corresponding to the ground states of neutral excitons ($X_{1s}^A$ and $X_{1s}^B$) and charged trions ($T^A$ and $T^B$) as well as up to eight additional spectral lines at energies above the $X_{1s}^B$ transition, which we attribute to the Rydberg series of excited states of $X^A$ and $X^B$. The intensities of the spectral features can be tuned by the gate and drain-source voltages. Using an effective-mass theory for excitons in two-dimensional systems we are able to accurately fit the measured spectral lines and unambiguously associate them with their corresponding Rydberg states.

**Introduction**

Two-dimensional transition metal dichalcogenides (2D-TMDs) are an excellent playground for studying and exploiting exciton physics. This family of materials presents unusually large exciton binding energies and lifetimes, even at room temperature and, in consequence, their optical and optoelectronic properties are largely dominated by excitonic transitions.[1]

Until recent years, research of exciton physics in 2D-TMDs has mainly relied on photoluminescence (PL) spectroscopy measurements. This technique and its variants (time-resolved PL,[2] PL excitation,[3] etc.) allowed to obtain detailed information on the properties of light-emitting exciton transitions, including exciton binding energy, [4,5] lifetime, [6,7] spin and valley polarization, [8,9] etc. However, PL spectroscopy relies on spontaneous radiative decay of excitonic states. Thus, if a competing non-radiative exciton relaxation mechanism is present, it can cause a dumping, or even complete disappearance of the corresponding PL peaks, even if they are allowed by optical selection rules.[10] In consequence, other techniques such as absorption spectroscopy, [11–15] or electroluminescence spectroscopy[16] have increasingly gained popularity for investigating excitonic states not accessible by PL.

Photocurrent spectroscopy (PCS)[17,18] provides a simple, powerful, and yet largely underused complementary approach for studying excitonic transitions in 2D-TMDs. For this technique, the sample at study is exposed to monochromatic light and the light-induced change in conductivity (photoconductivity) is registered as a function of the illumination wavelength. Since the detection mechanism does not require for excitons to decay radiatively via spontaneous emission, PCS allows detecting exciton transitions regardless of the presence of dominant non-radiative relaxation mechanisms. Further, since this technique is sensitive to the electric charge, it is especially suited to study the transport properties and ionization mechanisms of photogenerated excitons, which are crucial for the development of exciton-based optoelectronics. Up to date, however, only few attempts have been made to use PCS for exciton characterization in 2D materials,[18,19] and measurements at cryogenic temperature and with high-quality samples are still missing in literature.

Here we investigate the excitonic properties of a monolayer (1L) $MoS_2$ phototransistor by PCS at cryogenic temperature ($T$ = 5 K). Our measurements allow us to fully resolve and identify the 1L-$MoS_2$ neutral exciton states $X^A$ and $X^B$, originating from the two spin-split band-edge optical transitions at the K point of the reciprocal lattice, as well as their associated charged trion states, $T^A$ and $T^B$. Owing to the excellent quality of the fabricated device, in combination with the use of cryogenic temperature, we observe remarkably sharp exciton transitions, with bandwidths as low as 8 meV (full width at half maximum; FWHM), roughly one order of magnitude lower than in earlier PCS measurements,[18,19] and comparable to the bandwidths observed in low-temperature PL experiments for high-quality 1L-$MoS_2$.[20]

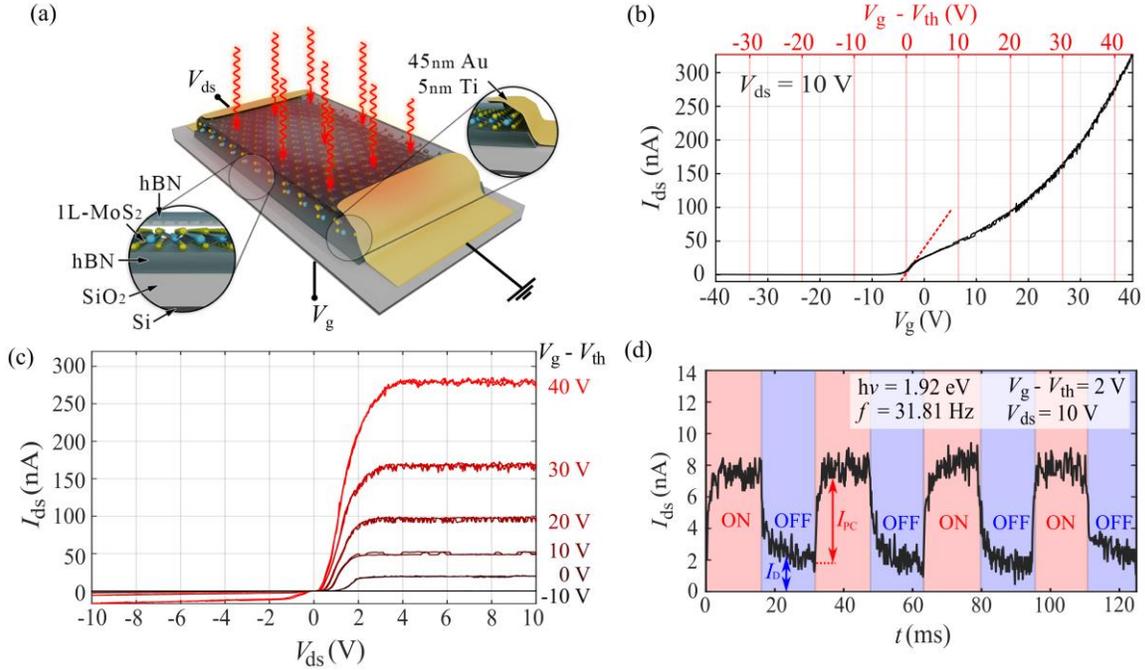

**Figure 1.** Measurement geometry and optoelectronic response of the 1L-MoS$_2$ phototransistor. (a) Schematic drawing of the fully hexagonal boron nitride (h-BN) encapsulated 1L-MoS$_2$ phototransistor with edge contacts. The insets at the top-right and bottom-left show in detail the geometry of the contacts and the 2D channel, respectively. For optoelectronic measurements, the entire device is exposed to homogeneous monochromatic illumination. (b) Gate transfer characteristic for $V_{sd}$ = 10 V, showing a clear n-type behaviour. The threshold gate voltage, $V_{th}$ = - 4 V, is estimated extrapolating the linear region of $I_{SD}$ (dashed red line). (c) I-V characteristics of the device measured for different gate voltages. (d) Source-drain current of the device for $V_{sd}$ = 10 V and $V_g - V_{th}$ = 2V. When the light excitation (hv = 1.92 eV) is turned on, the drain-source current increases by $I_{PC}$ = 6 nA.

By applying a gate voltage to increase the charge carrier density in the semiconductor channel we are able to tune the relative intensity of the exciton and trion transitions. This is observed not only for the $X^A$ and $T^A$ transitions – already reported in PL measurements – but also for the $X^B$ and $T^B$ features, not shown before. Moreover, we find that a drain-source voltage can also be used to modulate the relative intensity of exciton and trion spectral features in a similar way (shown in Supplementary Note 3).

Besides the four mentioned peaks, the measured PC spectra also show eight additional features at energies above the $X^B$ transition, which we attribute to the Rydberg series of excited states of $X^A$ and $X^B$.[3,15,21] Using a 2D effective-mass Hamiltonian with a non-hydrogenic Keldysh potential we are able to accurately reproduce the observed spectral features and unambiguously determine their origin.

## Results

*Device fabrication and electrical characterization*

Figure 1a schematically depicts the geometry of the studied 1L-MoS$_2$ phototransistor. A full description of the device fabrication process and the identification of monolayer flakes is provided in Supplementary Notes 1 and 2, respectively. The 2D channel is fully encapsulated between top and bottom multilayer hexagonal boron nitride (h-BN) flakes in order to better preserve its intrinsic properties.[22] The Ti/Au electrodes follow an edge-contact geometry, as shown at the top-right inset of Figure 1a and described in Supplementary Note 1.

We start our measurements by characterizing the electrical response of the device. Unless otherwise stated, all measurements reported below were performed in vacuum and at $T = 5$ K. Figure 1b shows a transfer curve of the monolayer MoS$_2$ phototransistor, obtained using the Si substrate as back gate. The curve shows a clear n-type behaviour, and the semiconductor channel conductivity increases as the back-gate voltage becomes larger than the threshold voltage ($V_{th}$ = -9 V).

Figure 1c shows two-terminal *I-V* curves of the device at different gate voltages. The curves present a back-to-back diode-like behaviour due to the presence of Schottky barriers at the contacts.[23,24] The different saturation currents for positive and negative voltages are caused by an asymmetry in the Schottky barrier heights. At low temperature ($T = 5$K), thermionic transport can be neglected as a mechanism for conduction and thus, the source-drain current $I_{sd}$ is mainly generated by tunnelling through the Schottky barriers.

Next, in order to characterize the 1L-MoS$_2$ photoresponse, we expose the whole device to a homogeneous monochromatic light source. Figure 1d shows the time-dependent source-drain current $I_{sd}$, measured while the optical excitation is turned on and off at a frequency $f = 31.81$ Hz, while applying a constant $V_{sd} = 10$ V and $V_g - V_{th} = 2$ V. The illumination energy is fixed at h$\nu$ = 1.92 eV, in resonance with the X$^A$ exciton (as shown below). When the light is turned on, $I_{ds}$ increases from its value in the dark, $I_D$, by an amount $I_{PC}$ due to photoconductivity.

In 2D-TMD phototransistors, photoconductivity can emerge from two main mechanisms:[18,25–29] photoconductive effect, where light-induced formation of electron–hole pairs leads to an increased charge carrier density and electrical conductivity; and photovoltaic effect, where the light-induced filling or depletion of localized states causes a shift of the Fermi energy. When the characteristic relaxation times for these localized states are very long, photovoltaic effects are observed as photodoping, rather than as photoconductivity, and the Fermi energy shift remains for a long time, or even permanently, after the optical excitation is removed.[30]

As we further discuss below and Supplementary Note 7, we believe that in our case the observed photoconductivity is mainly dominated by photovoltaic effects.[25] It is worth remarking that, due to the presence of h-BN layers in contact with the 1L-MoS$_2$ channel, the device is also affected by photodoping,[30] which can result in large and persistent shifts of $V_{th}$ after exposure to light. We find that this effect can be prevented to a large extent by carrying out the measurements at a low illumination power density, below 2.5 mW cm$^{-2}$. For such densities the light-induced shift of $V_{th}$ during the spectrum acquisition process remains lower than 0.1 V. To further avoid inconsistencies in the measurements due to shifts of $V_{th}$ all the spectral data is presented as a function of $V_g$-$V_{th}$.

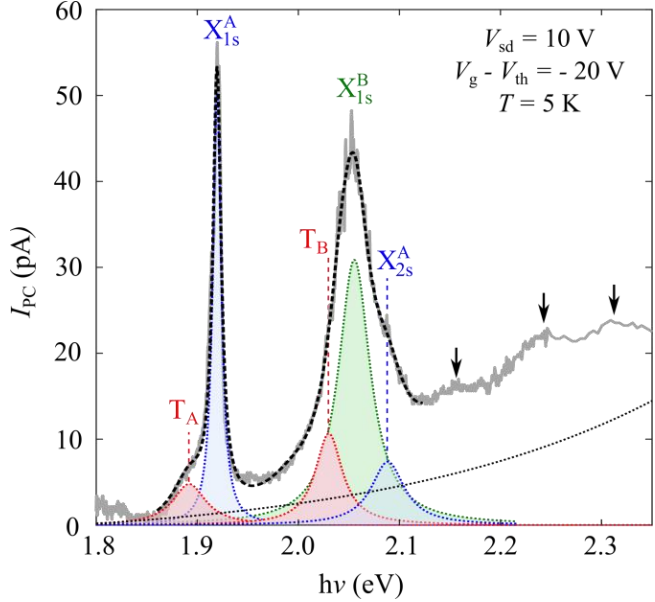

**Figure 2.** Photocurrent spectrum of the 1L-MoS$_2$ device (gray line) and fitting to a multi-peak Lorentzian plus an exponential background (black dashed line). The blue, green and red dotted lines are the single Lorentzian peaks corresponding to the five peaks described in the text, and the grey dotted line is the exponential background, accounting for the direct interband absorption edge. Downward arrows indicate additional spectral features occurring at energies above X$_{2s}^A$, which we attribute to Rydberg excited states of X$^A$ and X$^B$.

*Photocurrent spectra*

Figure 2 shows a typical low-temperature PC spectrum of the 1L-MoS$_2$ phototransistor, acquired at $T = 5$K, $V_{sd} = 10$ V and $V_g - V_{th} = -20$ V. To improve the signal-to-noise ratio, the measurement is performed while switching the illumination on and off at a fixed frequency of 31.81 Hz and the PC is registered using a lock-in amplifier. The PC spectrum is obtained by repeating this measurement while scanning the illumination wavelength in steps of 0.1 nm. A detailed description of the PCS setup can be found in Supplementary Note 4.

The resulting spectrum presents two main peaks at 1.918 eV and 2.060 eV, that we associate to the A and B excitonic ground states (X$_{1s}^A$ and X$_{1s}^B$) of 1L-MoS$_2$, with an energy splitting of 142 meV. We find that the linewidth of the observed peaks is at its lowest for gate voltages $V_g$ well below $V_{th}$. In these conditions the X$^A$ peak bandwidth is found to be as low as 8 meV (FWHM), roughly one order of magnitude lower than in earlier reports, and comparable with the typical A-exciton bandwidths for low-temperature PL spectroscopy in h-BN encapsulated 1L-MoS$_2$.[20] The observed narrow bandwidth confirms the high quality of the 2D semiconductor channel.

It is worth remarking that, differently from optical spectroscopy, PC spectral features can only emerge if a certain optical transition results in a change in photoconductivity. Thus, charge-neutral excitons can only be observed by this technique in presence of an efficient exciton dissociation mechanism. In few-layer TMD phototransistors, exciton dissociation processes are typically largely enhanced in the vicinity of metal-semiconductor contacts, due to the large electric fields present near the interfaces.[31–33] Thus, we expect that the observed PC is mainly produced in these regions.

For the spectral range between 1.85 eV and 2.15 eV the experimental PC spectrum can be very accurately fit by a multi-peak Lorentzian plus an exponential background, which accounts for the Urbach tail of the direct interband absorption edge. We find that the spectral profile is best reproduced by a quintuple Lorentzian function, with two main peaks centered at the energies of the $X_{1s}^A$ and $X_{1s}^B$ transitions discussed above plus three smaller peaks at 1.892 eV, 2.035 eV and 2.090 eV. We attribute the first two of these smaller features to the A and B trion states, $T^A$ and $T^B$, expected to occur at energies 20-30 meV below $X_{1s}^A$ and $X_{1s}^B$.[34] Photogenerated trions can be either positively or negatively charged depending on the nature of the constituent charge carriers, however, in our sample the MoS$_2$ channel is strongly n-doped and, thus, we expect that the observed $T^A$ and $T^B$ features mainly account for negatively charged trions. While here we discuss our results in terms of excitons and trions, it is worth noting that recent works have proposed an alternative description for X and T transitions as originated from interactions between excitons and the Fermi sea of excess carriers (either electrons or holes).[35,36] Under this picture, the spectral features typically assigned to neutral excitons and charged trions correspond to the repulsive and attractive exciton-polaron branches respectively.

The feature at 2.090 eV is tentatively assigned to the first excited Rydberg state of the A exciton, $X_{2s}^A$, recently observed in 1L-MoS$_2$ by low-temperature micro-reflectance and transmission spectroscopy measurements.[15,21] In fact, we also observe additional features lying at energies above 2.1 eV, indicated by arrows in Figure 2, which we associate to the Rydberg series of excited states of $X^A$ and $X^B$, as further discussed in below.

Although small, the three features assigned to $T^A$, $T^B$ and $X_{2s}^A$ are consistently reproduced in multiple spectra acquired at different bias voltages (see Supplementary Note 3). Furthermore, as we show below, these features become much more prominent when $V_g$ is increased to bring the MoS$_2$ Fermi energy above the edge of the conduction band.

*Gate modulation of photoconductivity and spectral features*

Figure 3a shows the gate dependence of $I_{PC}$ for illumination with h$\nu$ = 1.92 eV, on resonance with the $X^A$ transition. We find that the measured photocurrent becomes maximal for gate voltages close to the threshold voltage of the 1L-MoS$_2$ channel, i.e. when the Fermi energy approaches the edge of the conduction band, and largely decreases for sub-threshold voltages. This result is somewhat counterintuitive since for $V_g < V_{th}$ the electronic states at the conduction band should be completely depleted and thus, the probability of interband exciton absorption should be maximal. Indeed, for absorption spectroscopy experiments, the signal maximizes for Fermi energies below the edge of the conduction band, and decreases for larger gate voltages due to Pauli blockade.[37] For PC spectroscopy, however, the situation is more complex, as optical transitions will only be observed if they lead to a change in the device conductivity, either by a photoconductive effect or by a photovoltaic effect.

The observed decrease of PC for gate voltages below $V_{th}$ suggests that the main mechanism for photoresponse in our 1L-MoS$_2$ device is a photovoltaic effect similar to the one described by Furchi et al.[25] For this effect, upon optical excitation, photoexcited carriers decay into localized states within the 1L-MoS$_2$ bandgap, resulting in a shift of the Fermi energy. Differently from the h-BN induced photodoping effects mentioned above,[30] the relaxation time for charge carriers in these impurities is very short, and thus the effect manifests as an increase of conductivity while the device is exposed to light. A characteristic signature of the photovoltaic effect is that the resulting photocurrent is proportional to the transconductance $G = dI_{ds}/dV_g$ of the semiconductor channel. As shown in Figure 3a, the transconductance of the 1L-MoS$_2$ device, measured in the dark at $V_{sd} = 5$ V, markedly resembles the gate voltage dependence of $I_{PC}$, in consistence with the proposed mechanism for photoconductivity.

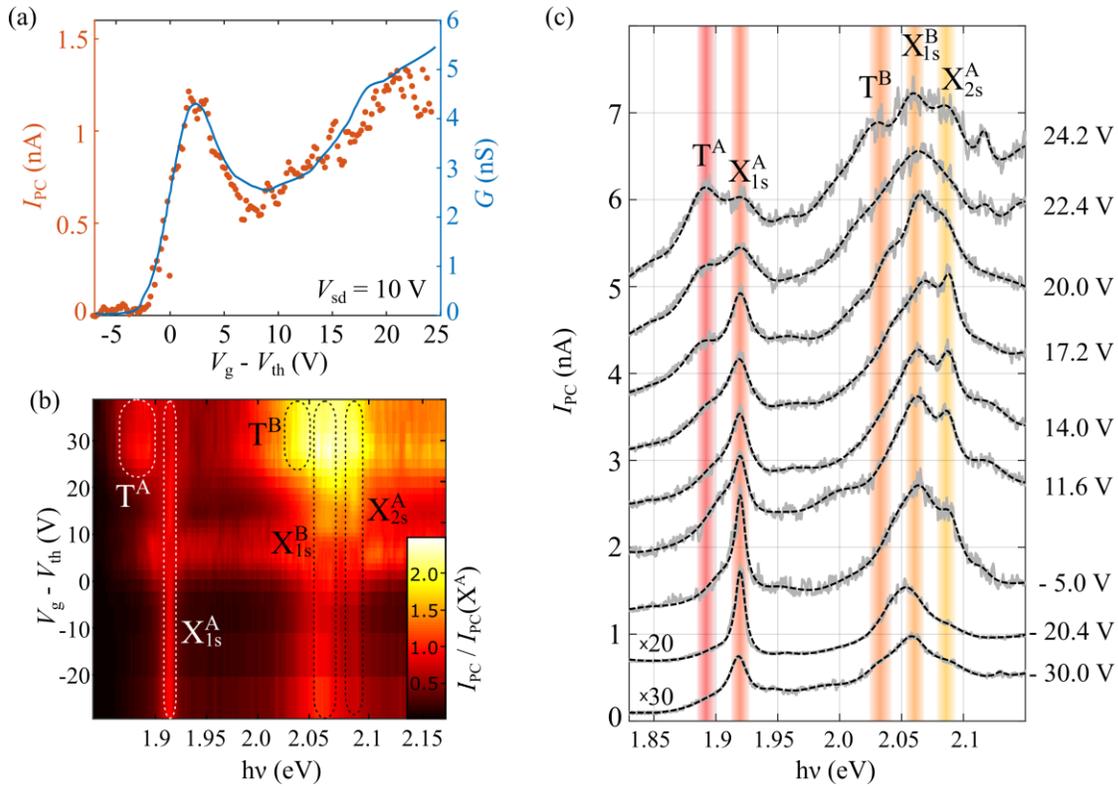

**Figure 3.** Gate dependence of the observed photocurrent. (a) Tranconductance measured in dark (blue line, right axis) and gate-dependent photocurrent measured at source-drain voltage $V_{sd}$ = 10 V for illumination on resonance with the X$^A$ exciton transition (orange dots, left axis). Error bars are smaller than the data points. (b) Colormap of the measured photocurrent as a function of the excitation energy and the applied gate voltage. For each individual gate voltage the photocurrent data has been normalized to the value measured at $h\nu$ = 1.92 eV, on resonance with the X$^A$ transition. (c) Individual photocurrent spectra acquired at different values of gate voltage $V_g$, with $V_g - V_{th}$ ranging from -30 to 24.2 V (gray lines). The dashed lines are fits of the measured spectra to a multi-peak Lorentzian plus an exponential background. For clarity, the spectra have been shifted vertically in steps of 0.6 nA.

Next we address the effect of the gate voltage on the PC spectral features. For simplicity we restrict our discussion here to the five spectral lines discussed above (higher-energy spectral features will be discussed in the next section). Figure 3b shows a color map of the photocurrent as a function of the excitation energy h$\nu$ and the gate voltage $V_g - V_{th}$. For each value of $V_g - V_{th}$ the photocurrent data has been normalized to the value measured at h$\nu = 1.92$ eV, on resonance with the X$^A$ transition. The five excitonic features discussed above are also apparent here, and their relative intensities are largely modulated by the gate voltage, as more clearly observed in the individual spectra shown in Figure 3c and discussed below. A similar modulation of spectral features is also observed when tuning the drain-source voltage, as described in Supplementary Note 3.

When $V_g$ is increased, the trion transition T$^A$ becomes progressively larger, even becoming more prominent than X$^A$ for $V_g - V_{th} = 24.2$ V. A similar gate voltage modulation of the X$^A_{1s}$ and T$^A$ spectral features in 1L-TMDs has been reported in literature for photoluminescence,[38,39] electroluminescence,[16] and absorption spectroscopy[37] measurements. Typically, the T$^A$ spectral feature becomes more prominent when the Fermi energy is set above the conduction band edge, since in this situation excess electrons in the semiconductor channel can efficiently bind with photoexcited electron-hole pairs.[37] Similar to T$^A$, we find that the T$^B$ transition also becomes more prominent as the electron density in the conduction band is increased by the gate voltage.

As shown in earlier literature for monolayer TMDs, applying a gate voltage $V_g > V_{th}$ also results in an increased energy splitting between the X$^A$ and T$^A$ absorption peaks.[37,40] While this gate modulation should be visible in the PC spectra, here we only observe it very weakly, as we do not reach sufficiently large doping levels to fully resolve this effect (see Supplementary Note 10).

Finally, as $V_g$ increases, we also observe a strengthening of the peak associated to the X$^A_{2s}$ excited state, which even becomes larger than X$^B_{1s}$ for a certain gate voltage range. As further discussed in next section and in Supplementary Note 6, we observe a similar gate modulation for excited Rydberg states of X$^A$ and X$^B$ laying at higher energies.

*Rydberg series*

We now turn to the study of the different spectral features observed for energies above X$^B_{1s}$. Figure 4a shows the photocurrent spectrum of the 1L-MoS$_2$ device at $T = 5$ K, $V_{sd} = 10$ V and $V_g - V_{th} = 17.2$ V. In addition to the peaks associated to X$^A_{1s}$, X$^B_{1s}$, T$^A$ and T$^B$, we observe eight additional peaks, which we tentatively assign to the Rydberg series of excited states of X$^A$ and X$^B$, as labelled in the figure and enlisted in Table 1. Although some of these peaks are relatively weak compared to the X$_{1s}^A$ and X$_{1s}^B$ features, they consistently appear in spectra acquired for different gate and drain-source voltages (see Supplementary Note 6).

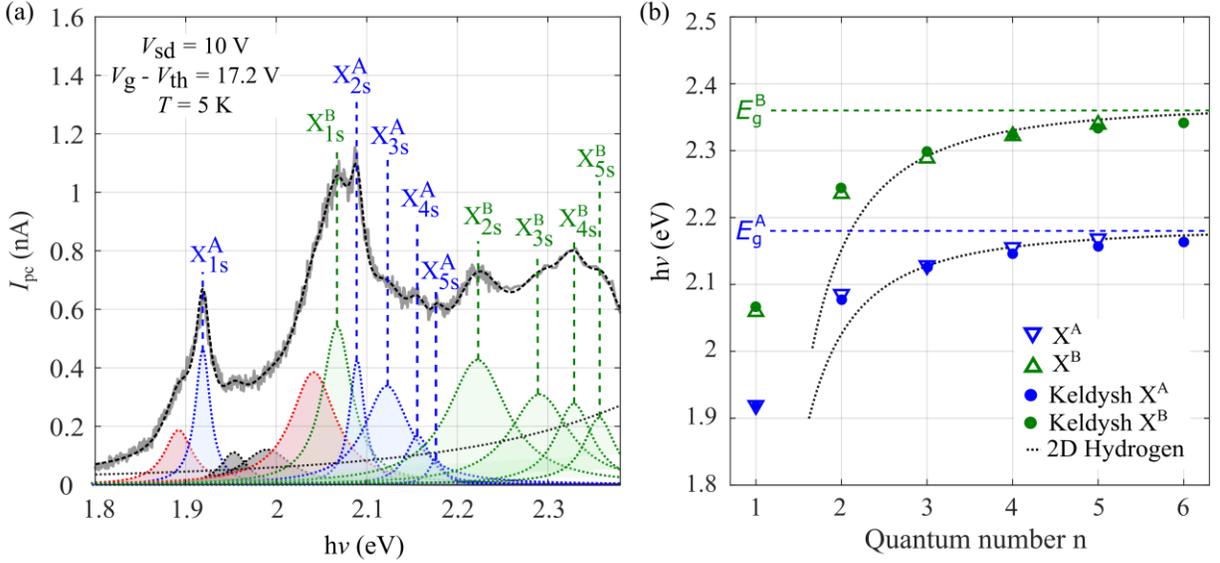

**Figure 4.** Photocurrent (PC) spectroscopy of excited Rydberg states. (a) PC spectrum for the 1L-MoS$_2$ device (gray, solid line) and fit to a multiple-peak Lorentzian function (black, dashed line). The individual Lorentzian functions corresponding to Rydberg series of excitons X$^A$ and X$^B$ are shown in blue and green respectively. (b) Experimental (empty triangles) and theoretical (filled circles, solid lines) transition energies for the exciton states. Theoretical values are obtained by fitting to the effective mass model described in the main text. The fits of the $n > 2$ states to a 2D hydrogen model (black dashed lines) are shown for comparison. Dashed horizontal lines indicate the theoretically calculated values for the quasiparticle bandgaps $E_g^A = 2.18$ eV and $E_g^B = 2.36$ eV. For the experimental data acquired at different gate voltages the measured peak positions fluctuate by, at most, 6 meV.

It is worth noting that, while $s$-type transitions are predicted to be dipole-allowed, and therefore visible in the linear photoconductivity spectra,[41] $p$ and $d$ excited states do not directly couple to light. Thus, in the following we restrict our discussion to $s$ states only.

**Table 1.** Position of exciton transitions in the 1L-MoS$_2$ PC spectrum. The energy position of the exciton transitions is extracted from the photocurrent spectrum in Figure 4 by multi-Lorentzian fitting. For spectra acquired at different gate voltages we observe small fluctuations in the peak positions of, at most, ± 6 meV.

|  | T$^A$ | X$_{1s}^A$ | T$^B$ | X$_{1s}^B$ | X$_{2s}^A$ | X$_{3s}^A$ | X$_{4s}^A$ | X$_{5s}^A$ | X$_{2s}^B$ | X$_{3s}^B$ | X$_{4s}^B$ | X$_{5s}^B$ |
|---|---|---|---|---|---|---|---|---|---|---|---|---|
| h$\nu$ (eV) | 1.892 | 1.919 | 2.041 | 2.067 | 2.089 | 2.127 | 2.161 | 2.172 | 2.231 | 2.298 | 2.329 | 2.349 |

Figure 4b shows the spectral position of the observed peaks and their tentatively associated quantum number *n*. In order to confirm unambiguously the spectral assignments of the peaks we fit their spectral positions using an effective-mass theory for excitons in 2D-TMDs.[42]

We solve numerically the effective-mass Schrödinger equation for the radially symmetric exciton states *ns*, with energies $E_n$:

$$H\psi_{ns}(r) = (E_n - E_g)\psi_{ns}(r). \tag{1}$$

Here $E_g$ is the quasiparticle bandgap and $H$ is the Hamiltonian for the relative coordinate $r = |r_e - r_h|$, namely

$$H = -(\hbar^2/2\mu)\, d^2/dr^2 + V(r). \tag{2}$$

For the A and B exciton effective masses we take $\mu_A = 0.27\, m_0$[21] and $\mu_B = 0.28\, m_0$,[42] respectively, being $m_0$ the free electron mass. In thin semiconductor layers, the electron-hole interaction $V(r)$ cannot be simply modelled as a Coulomb potential because it is largely affected by nonlocal screening from the embedding medium. Instead, the screened electron-hole interaction is accurately described by the Keldysh potential [43]

$$V(r) = -(\pi e^2/2r_0)[H_0(\kappa r/r_0) - Y_0(\kappa r/r_0)]. \tag{3}$$

Here $H_0$ and $Y_0$ are zero-order Struve and Bessel functions and $\kappa$ is the dielectric constant of the embedding medium (h-BN in our case). The parameter $r_0$ is related to the screening length due to the 2D polarizability of the 1L-MoS$_2$. For the discussion below, it is worth mentioning that the Keldysh potential approaches the Coulomb potential $V(r) \approx -e^2/\kappa r$ at large distance ($r \gg r_0/\kappa$) but diverges logarithmically at short distance ($r \ll r_0/\kappa$).

Thus, we are left with three fitting parameters for each excitonic Rydberg series, namely $E_g$, $\kappa$ and $r_0$. A first estimation of the quasiparticle bandgap $E_g$ can be deduced from the spectral position of the highly excited states ($n > 2$). For these states, the radius of the exciton lies in the region where $V(r) \approx -e^2/\kappa r$ and the corresponding energy levels can be nicely fitted by 2D hydrogenic Rydberg series

$$E_n = E_g - \frac{\mu e^4/(2\hbar^2\kappa^2)}{(n - 1/2)^2}. \tag{4}$$

From the fitting we get $E_g^A = 2.203$ eV and $E_g^B = 2.387$ eV. We then use these values as an initial guess and perform a more accurate fit using the Keldysh potential and numerically solving the corresponding Schrödinger equation (discussed in Supplementary Note 5).

The fit reproduces with great accuracy the energies of the experimentally observed spectral features as shown in Figure 4b, and allows us to estimate the optoelectronic parameters of 1L-MoS$_2$. The results of the fit are summarized in Table 2. For the X$^A$ Rydberg series we obtain a quasiparticle bandgap of 2.18 eV. Very recently Goryca et al. used absorption spectroscopy at high magnetic fields to estimate the quasiparticle bandgap of h-BN encapsulated MoS$_2$. There, they obtained $E_g^A = 2.16$ eV, in good agreement with the value reported here.[21] For the dielectric screening parameters we get $r_0^A = 3.05$ nm and $\kappa = 4.4$, also similar to earlier results in h-BN encapsulated samples.[15,21] Finally, we can estimate the binding energy of X$^A$ as $E_b^A = E_g^A - E_{1s}^A$. We get $E_b^A = 261$ meV, again in reasonable agreement with absorption and micro-reflectance spectroscopy measurements in h-BN encapsulated 1L-MoS$_2$.[15,21]

For the X$^B$ Rydberg series information is scarce in literature and there are no earlier results for h-BN encapsulated samples. Hill et al.[3] reported the observation of X$^B$ excited states in the room-temperature PL emission spectrum of 1L-MoS$_2$. There, they estimated a quasiparticle bandgap of $E_g^B = 2.47$ eV and an exciton binding energy of about 440 meV. From our measurements on h-BN encapsulated 1L-MoS$_2$ we get $E_g^B = 2.36$ eV and an exciton binding energy $E_b^B = 290$ meV, about 100 meV lower.

Finally, we use the obtained values of $E_g^A$ and $E_g^B$ to give an estimation of the MoS$_2$ valence band splitting $\Delta E_{VB}$. Assuming a conduction band splitting of $\Delta E_{CB} \approx 15$ meV [44,45] we get $\Delta E_{VB} \approx E_g^B - E_g^A - \Delta E_{CB} = 165$ meV, slightly above the value recently measured by ARPES for epitaxial 1L-MoS$_2$ on gold ($\Delta E_{VB} = 145 \pm 4$ meV).[46] Note that, using the value for $E_g^B$ given in ref. 3 would yield an even larger $\Delta E_{VB} \approx 275$ meV.

**Table 2.** Fundamental optoelectronic material parameters of monolayer MoS$_2$.
Experimentally determined values of exciton binding energies ($E_b^A$ and $E_b^B$), quasiparticle bandgaps ($E_g^A$ and $E_g^B$) and dielectric screening parameters ($r_0^A$, $r_0^B$ and $\kappa$), for A and B excitons in monolayer MoS$_2$. For our work, typical error bars for exciton binding energies and quasiparticle bandgaps are ± 5 meV. Typical error bars for $r_0^A$ and $r_0^B$ are ± 0.1 nm. For kappa we get error bars of ± 0.1.

| | $E_b^A$ (meV) | $E_b^B$ (meV) | $E_g^A$ (eV) | $E_g^B$ (eV) | $r_0^A$ (nm) | $r_0^B$ (nm) | $\kappa$ |
|---|---|---|---|---|---|---|---|
| Goryca 2019 [21] | 221 | | 2.16 | | 3.40 | | 4.45 |
| Robert 2018 [15] | 222 | | | | 2.95 | | |
| Hill 2015 [3] | | 440 | | 2.47 | | | |
| Our work | 261 | 290 | 2.18 | 2.36 | 3.05 | 2.70 | 4.4 |

**Discussion**

As we showed, low-temperature PCS allows us to observe very sharp excitonic spectral features, with linewidths as low as 8 meV (FWHM). While similar bandwidths for exciton features can also be achieved by PL or optical spectroscopy techniques, this typically requires the use of a microscope objective to concentrate the beam on a small area of the sample (in the order of 1 μm$^2$) to prevent peak broadening due to sample inhomogeneities. In our case however, the area of the sample that contributes to the observed spectrum is delimited by the spacing between the drain and source contacts (see Supplementary Note 4). This allows exposing the whole sample to light without losing spectral resolution, largely simplifying the experimental setup, as well as the procedure for optical alignment.

Using PCS we were able to address spectral features typically difficult to observe in PL due to their relatively low PL emission intensity. Owing to this fact, we could observe the electric field modulation of not only the X$^A$ and T$^A$ transitions, already described in literature for PL, but also of the X$^B$ and T$^B$ transitions, which to our knowledge was not reported in literature so far.

Finally, we were also able to clearly observe the excited Rydberg states of X$^A$ and X$^B$, up to n=5. Similar spectral features have been also observed in previous experimental works by PL spectroscopy, PL emission and micro reflectance, [3,15,47] as well as predicted in theoretical studies. [34,41,42,48] However, earlier experiments only revealed excited states corresponding to the Rydberg series of either X$^A$ or X$^B$, but never both combined, making difficult to unequivocally label the observed spectral features. The PC spectra presented here, on the other hand, cannot be explained by considering only one series of Rydberg states, but require accounting for excited states of both X$^A$ and X$^B$. This strongly constrains the possible spectral assignments. The peak fittings also allow us to extract optoelectronic material parameters for encapsulated 1L-MoS$_2$ that agree well with theoretical and experimental literature, further supporting the proposed peak assignments. Thanks to the simultaneous measurement of X$^A$ and X$^B$, we could also estimate the valence band splitting of 1L-MoS$_2$, not calculated before by this method, obtaining $\Delta E_{VB} \approx E_g^B - E_g^A = 165$ meV. Earlier theoretical works give similar values for $\Delta E_{VB}$, within 140-170 meV.[49,50] A recent ARPES measurement [46] for epitaxial 1L-MoS$_2$ on gold gave a slightly lower value, $\Delta E_{VB} = 145 \pm 4$ meV. However, this lower value could be explained by the effect of enhanced screening by the gold substrate.[34]

In all, this work demonstrates low-temperature photocurrent spectroscopy as a simple and powerful technique with great potential for the study of optoelectronics and exciton physics in two-dimensional materials. In particular, the diversity of excitonic features identified here suggests that low-temperature PC spectroscopy could be favourable to other spectral techniques for observing excited excitonic states and exciton complexes with relatively small binding energies, since the larger exciton ionization rate for these states results in a strengthening of their associated spectral features. We thus expect that, in the near future, this characterization technique will progressively become more popular among the 2D optoelectronics community.

**Acknowledgments**

We acknowledge financial support from the Agencia Estatal de Investigación of Spain (Grants MAT2016-75955, PID2019-106820RB and RTI2018-097180-B-100) and the Junta de Castilla y León (Grant SA256P18), including funding by ERDF/FEDER. J.Q. acknowledges his research contract funded by Junta de Castilla y León and FEDER funds. We are also thankful to Mercedes Velázquez for her help with the photoluminescence and Raman characterization.

**Author Contributions**

E.Diez and J.Q. conceived and supervised the research, D.V., A.M.-R., Y.M.M., and J.Q developed and tested the experimental setup for photocurrent spectroscopy. V.C., J.S.-S. and J.Q. fabricated and characterized the monolayer $MoS_2$ phototransistors, D.V. and J.Q carried out the electronic, optoelectronic, and spectral measurements and data analysis, D.V., E.Díaz and F.D.-A. performed the theoretical analysis of Rydberg states. The article was written through contribution of all the authors, coordinated by J.Q.

**Data Availability**

The data that support the findings of this study are available from the corresponding author upon request.

**References**

(1) Splendiani, A.; Sun, L.; Zhang, Y.; Li, T.; Kim, J.; Chim, C.-Y.; Galli, G.; Wang, F. Emerging Photoluminescence in Monolayer MoS2. *Nano Lett.* **2010**, *10*, 1271–1275.

(2) Wang, G.; Palleau, E.; Amand, T.; Tongay, S.; Marie, X.; Urbaszek, B. Polarization and Time-Resolved Photoluminescence Spectroscopy of Excitons in MoSe2 Monolayers. *Appl. Phys. Lett.* **2015**, *106*.

(3) Hill, H. M.; Rigosi, A. F.; Roquelet, C.; Chernikov, A.; Berkelbach, T. C.; Reichman, D. R.; Hybertsen, M. S.; Brus, L. E.; Heinz, T. F. Observation of Excitonic Rydberg States in Monolayer MoS2 and WS2 by Photoluminescence Excitation Spectroscopy. *Nano Lett.* **2015**, *15*, 2992–2997.

(4) Ugeda, M. M.; Bradley, A. J.; Shi, S.; Jornada, F. H.; Zhang, Y.; Qiu, D. Y.; Ruan, W.; Mo, S.-K.; Hussain, Z.; Shen, Z.; *et al.* Giant Bandgap Renormalization and Excitonic Effects in a Monolayer Transition Metal Dichalcogenide Semiconductor. *Nat. Mater.* **2014**, *13*, 1091–1095.

(5) Chernikov, A.; Berkelbach, T. C.; Hill, H. M.; Rigosi, A.; Li, Y.; Aslan, O. B.; Reichman, D. R.; Hybertsen, M. S.; Heinz, T. F. Exciton Binding Energy and Nonhydrogenic Rydberg Series in Monolayer WS 2. *Phys. Rev. Lett.* **2014**, *076802*, 1–5.

(6) Robert, C.; Lagarde, D.; Cadiz, F.; Wang, G.; Lassagne, B.; Amand, T.; Balocchi, A.; Renucci, P.; Tongay, S.; Urbaszek, B.; *et al.* Exciton Radiative Lifetime in Transition Metal Dichalcogenide Monolayers. *Phys. Rev. B - Condens. Matter Mater. Phys.* **2016**, *93*, 1–10.

(7) Fang, H. H. H.; Han, B.; Robert, C.; Semina, M. A. A.; Lagarde, D.; Courtade, E.; Watanabe, K.; Amand, T.; Urbaszek, B.; Glazov, M. M. M.; *et al.* Control of the Exciton Radiative Lifetime in van Der Waals Heterostructures. *Phys. Rev. Lett.* **2019**, *123*, 67401.

(8) Xiao, D.; Liu, G. Bin; Feng, W.; Xu, X.; Yao, W. Coupled Spin and Valley Physics in Monolayers of MoS 2 and Other Group-VI Dichalcogenides. *Phys. Rev. Lett.* **2012**, *108*, 196802.

(9) Zeng, H.; Dai, J.; Yao, W.; Xiao, D.; Cui, X. Valley Polarization in MoS2 Monolayers by Optical Pumping. *Nat. Nanotechnol.* **2012**, *7*, 490–493.

(10) Molas, M. R.; Faugeras, C.; Slobodeniuk, A. O.; Nogajewski, K.; Bartos, M.; Basko, D. M.; Potemski, M. Brightening of Dark Excitons in Monolayers of Semiconducting Transition Metal Dichalcogenides. *2D Mater.* **2016**, *4*, 21003.

(11) Castellanos-Gomez, A.; Quereda, J.; van der Meulen, H. P.; Agraït, N.; Rubio-Bollinger, G. Spatially Resolved Optical Absorption Spectroscopy of Single- and Few-Layer MoS2 by Hyperspectral Imaging. *Nanotechnology* **2016**, *27*, 1–16.

(12) He, K.; Poole, C.; Mak, K. F.; Shan, J. Experimental Demonstration of Continuous Electronic Structure Tuning via Strain in Atomically Thin MoS2. *Nano Lett.* **2013**, *13*, 2931–2936.

(13) Frisenda, R.; Niu, Y.; Gant, P.; Molina-Mendoza, A. J.; Schmidt, R.; Bratschitsch, R.; Liu, J.; Fu, L.; Dumcenco, D.; Kis, A.; *et al.* Micro-Reflectance and Transmittance Spectroscopy: A Versatile and Powerful Tool to Characterize 2D Materials. *J. Phys. D. Appl. Phys.* **2017**, *50*, 074002.

(14) Zhang, C.; Wang, H.; Chan, W.; Manolatou, C.; Rana, F. Absorption of Light by Excitons and Trions in Monolayers of Metal Dichalcogenide Mo S2: Experiments and Theory. *Phys. Rev. B - Condens. Matter Mater. Phys.* **2014**, *89*.

(15) Robert, C.; Semina, M. A.; Cadiz, F.; Manca, M.; Courtade, E.; Taniguchi, T.; Watanabe, K.; Cai, H.; Tongay, S.; Lassagne, B.; *et al.* Optical Spectroscopy of Excited Exciton States in MoS 2 Monolayers in van Der Waals Heterostructures. *Phys. Rev. Mater.* **2018**, *2*, 1–9.
(16) Paur, M.; Molina-Mendoza, A. J.; Bratschitsch, R.; Watanabe, K.; Taniguchi, T.; Mueller, T. Electroluminescence from Multi-Particle Exciton Complexes in Transition Metal Dichalcogenide Semiconductors. *Nat. Commun.* **2019**, *10*, 1709.
(17) Kam, K. K.; Parklnclon ', A. Detailed Photocurrent Spectroscopy of the Semiconducting Group V I Transition Metal Dichaicogenldes. *J. Phys. Chem* **1982**, *86*, 463–467.
(18) Quereda, J.; Ghiasi, T. S.; van Zwol, F. A.; van der Wal, C. H.; van Wees, B. J. Observation of Bright and Dark Exciton Transitions in Monolayer MoSe2 by Photocurrent Spectroscopy. *2D Mater.* **2018**, *5*, 015004.
(19) Klots, A. R.; Newaz, A. K. M.; Wang, B.; Prasai, D.; Krzyzanowska, H.; Lin, J.; Caudel, D.; Ghimire, N. J.; Yan, J.; Ivanov, B. L.; *et al.* Probing Excitonic States in Suspended Two-Dimensional Semiconductors by Photocurrent Spectroscopy. *Sci. Rep.* **2014**, *4*, 6608.
(20) Cadiz, F.; Courtade, E.; Robert, C.; Wang, G.; Shen, Y.; Cai, H.; Taniguchi, T.; Watanabe, K.; Carrere, H.; Lagarde, D.; *et al.* Excitonic Linewidth Approaching the Homogeneous Limit in MoS2-Based van Der Waals Heterostructures. *Phys. Rev. X* **2017**, *7*, 1–12.
(21) Goryca, M.; Li, J.; Stier, A. V; Taniguchi, T.; Watanabe, K.; Urbaszek, B.; Marie, X.; Crooker, S. A.; Courtade, E.; Shree, S.; *et al.* Revealing Exciton Masses and Dielectric Properties of Monolayer Semiconductors with High Magnetic Fields. *Nat. Commun.* **2019**, *10*, 4172.
(22) Das, T.; Ahn, J.-H. Development of Electronic Devices Based on Two-Dimensional Materials. *FlatChem* **2017**, *3*, 43–63.
(23) Chiquito, A. J.; Amorim, C. a; Berengue, O. M.; Araujo, L. S.; Bernardo, E. P.; Leite, E. R. Back-to-Back Schottky Diodes: The Generalization of the Diode Theory in Analysis and Extraction of Electrical Parameters of Nanodevices. *J. Phys. Condens. Matter* **2012**, *24*, 225303.
(24) Quereda, J.; Palacios, J. J.; Agräit, N.; Castellanos-Gomez, A.; Rubio-Bollinger, G. Strain Engineering of Schottky Barriers in Single- and Few-Layer MoS2 Vertical Devices. *2D Mater.* **2017**, *4*, 021006.
(25) Furchi, M. M.; Polyushkin, D. K.; Pospischil, A.; Mueller, T. Mechanisms of Photoconductivity in Atomically Thin MoS2. *Nano Lett.* **2014**, *14*, 22.
(26) Miller, B.; Parzinger, E.; Vernickel, A.; Holleitner, A. W.; Wurstbauer, U. Photogating of Mono- and Few-Layer MoS2. *Appl. Phys. Lett.* **2015**, *106*, 1–5.
(27) Island, J. O.; Blanter, S. I.; Buscema, M.; Van Der Zant, H. S. J. J.; Castellanos-Gomez, A. Gate Controlled Photocurrent Generation Mechanisms in High-Gain In2Se3 Phototransistors. *Nano Lett.* **2015**, *15*.
(28) Fang, H.; Hu, W. Photogating in Low Dimensional Photodetectors. *Adv. Sci.* **2017**, *4*.
(29) Huang, H.; Wang, J.; Hu, W.; Liao, L.; Wang, P.; Wang, X.; Gong, F.; Chen, Y.; Wu, G.; Luo, W.; *et al.* Highly Sensitive Visible to Infrared MoTe 2 Photodetectors Enhanced by the Photogating Effect. *Nanotechnology* **2016**, *27*, 445201.
(30) Quereda, J.; Ghiasi, T. S.; Van Der Wal, C. H.; Van Wees, B. J. Semiconductor Channel-Mediated Photodoping in h-BN Encapsulated Monolayer MoSe2 Phototransistors. *2D Mater.* **2019**, *6*.
(31) Pedersen, T. G.; Latini, S.; Thygesen, K. S.; Mera, H.; Nikolić, B. K. Exciton Ionization in Multilayer Transition-Metal Dichalcogenides. *New J. Phys.* **2016**, *18*.
(32) Massicotte, M.; Vialla, F.; Schmidt, P.; Lundeberg, M. B.; Latini, S.; Haastrup, S.; Danovich, M.; Davydovskaya, D.; Watanabe, K.; Taniguchi, T.; *et al.* Dissociation of Two-Dimensional Excitons in Monolayer WSe2. *Nat. Commun.* **2018**, *9*, 1–7.

(33) Kamban, H. C.; Pedersen, T. G. Field-Induced Dissociation of Two-Dimensional Excitons in Transition Metal Dichalcogenides. *Phys. Rev. B* **2019**, *100*, 1–8.

(34) Drüppel, M.; Deilmann, T.; Krüger, P.; Rohlfing, M. Diversity of Trion States and Substrate Effects in the Optical Properties of an MoS2 Monolayer. *Nat. Commun.* **2017**, *8*, 1–7.

(35) Sidler, M.; Back, P.; Cotlet, O.; Srivastava, A.; Fink, T.; Kroner, M.; Demler, E.; Imamoglu, A. Fermi Polaron-Polaritons in Charge-Tunable Atomically Thin Semiconductors. *Nat. Phys.* **2017**, *13*, 255–261.

(36) Efimkin, D. K.; MacDonald, A. H. Many-Body Theory of Trion Absorption Features in Two-Dimensional Semiconductors. *Phys. Rev. B* **2017**, *95*, 35417.

(37) Mak, K. F.; He, K.; Lee, C.; Lee, G. H.; Hone, J.; Heinz, T. F.; Shan, J. Tightly Bound Trions in Monolayer MoS 2. *Nat. Mater.* **2013**, *12*, 207–211.

(38) Yang, J.; Lü, T.; Myint, Y. W.; Pei, J.; Macdonald, D.; Zheng, J. C.; Lu, Y. Robust Excitons and Trions in Monolayer MoTe2. *ACS Nano* **2015**, *9*, 6603–6609.

(39) Ross, J. S.; Wu, S.; Yu, H.; Ghimire, N. J.; Jones, A. M.; Aivazian, G.; Yan, J.; Mandrus, D. G.; Xiao, D.; Yao, W.; *et al.* Electrical Control of Neutral and Charged Excitons in a Monolayer Semiconductor. *Nat. Commun.* **2013**, *4*, 1474.

(40) Chernikov, A.; Zande, A. M. Van Der; Hill, H. M.; Rigosi, A. F.; Velauthapillai, A.; Hone, J.; Heinz, T. F.; van der Zande, A. M.; Hill, H. M.; Rigosi, A. F.; *et al.* Electrical Tuning of Exciton Binding Energies in Monolayer WS2. *Phys. Rev. Lett.* **2015**, *126802*, 1–6.

(41) Berghäuser, G.; Malic, E. Analytical Approach to Excitonic Properties of MoS 2. *Phys. Rev. B - Condens. Matter Mater. Phys.* **2014**, *89*, 1–7.

(42) Berkelbach, T. C.; Hybertsen, M. S.; Reichman, D. R. Theory of Neutral and Charged Excitons in Monolayer Transition Metal Dichalcogenides. *Phys. Rev. B* **2013**, *88*, 45318.

(43) Keldysh, L. Coulomb Interaction in Thin Semiconductor and Semimetal Films. *Soviet Journal of Experimental and Theoretical Physics Letters*, 1979, *29*, 658.

(44) Robert, C.; Han, B.; Kapuscinski, P.; Delhomme, A.; Faugeras, C.; Amand, T.; Molas, M. R. Measurement of the Spin-Forbidden Dark Excitons in MoS 2 and MoSe 2 Monolayers.

(45) Pisoni, R.; Kormányos, A.; Brooks, M.; Lei, Z.; Back, P.; Eich, M.; Overweg, H.; Lee, Y.; Rickhaus, P.; Watanabe, K.; *et al.* Interactions and Magnetotransport through Spin-Valley Coupled Landau Levels in Monolayer MoS2. *Phys. Rev. Lett.* **2018**, *121*, 247701.

(46) Miwa, J. A.; Ulstrup, S.; Sørensen, S. G.; Dendzik, M.; Čabo, A. G.; Bianchi, M.; Lauritsen, J. V.; Hofmann, P.; Grubi, A.; Bianchi, M.; *et al.* Electronic Structure of Epitaxial Single-Layer MoS 2. *Phys. Rev. Lett.* **2015**, *114*, 46802.

(47) Pandey, J.; Soni, A. Unraveling Biexciton and Excitonic Excited States from Defect Bound States in Monolayer MoS 2. *Appl. Surf. Sci.* **2018**, *463*, 52–57.

(48) Wu, F.; Qu, F.; Macdonald, A. H. Exciton Band Structure of Monolayer MoS2. *Phys. Rev. B - Condens. Matter Mater. Phys.* **2015**, *91*, 1–9.

(49) Zhu, Z. Y.; Cheng, Y. C.; Schwingenschlögl, U.; Schwingenschl, U. Giant Spin-Orbit-Induced Spin Splitting in Two-Dimensional Transition-Metal Dichalcogenide Semiconductors. *Phys. Rev. B* **2011**, *84*, 153402.

(50) Ramasubramaniam, A. Large Excitonic Effects in Monolayers of Molybdenum and Tungsten Dichalcogenides. *Phys. Rev. B* **2012**, *86*, 115409.

Supplementary Information to: Excitons, trions and Rydberg states in monolayer MoS$_2$ revealed by low-temperature photocurrent spectroscopy.

*Daniel Vaquero[1], Vito Clericò[1], Juan Salvador-Sánchez[1], Adrián Martín-Ramos[1], Elena Díaz[2], Francisco Domínguez-Adame[2], Yahya M. Meziani[1], Enrique Diez[1] and Jorge Quereda[1]\**

[1] Nanotechnology Group, USAL–Nanolab, Universidad de Salamanca, E-37008 Salamanca, Spain
[2] GISC, Departamento de Física de Materiales, Universidad Complutense, E-28040 Madrid, Spain

\* e-mail: j.quereda@usal.es

**Table of contents**

**Supplementary Note 1: Device fabrication and contact geometry**

Figure S1 summarizes the main steps for the device fabrication. The process starts with the stacking of the heterostructure of single layer (1L) MoS$_2$ completely encapsulated in hexagonal boron nitride (h-BN), using a dry-transfer method similar to the polypropylene carbonate (PPC) method for van der Waals heterostructures [1]. We first exfoliate MoS$_2$ and h-BN flakes by the standard scotch-tape method and transfer the flakes onto SiO$_2$ substrates. Then, we inspect the substrates through optical microscope (Figure S1a), identify the 1L-MoS$_2$ flakes by their optical contrast and confirm their thickness by micro-Raman spectroscopy as further detailed in section S2. We also use optical miscroscopy to identify and select two h-BN flakes with a thickness of 15-20 nm for the top layer h-BN and 25-30 nm for the bottom layer one.

Next, we start the stacking process by transferring the top h-BN layer onto the MoS$_2$ flake (Figure S1b). Subsequently we clean the substrate containing the top h-BN/MoS$_2$ heterostructure with anisole, acetone and isopropanol (IPA) for few minutes. Both flakes are then picked up together using a PPC film, as described in ref. 1, and transferred onto the bottom layer h-BN, previously cleaned with acetone and IPA and annealed at 380 ºC for 15 minutes in an Argon atmosphere. Once the entire heterostructure is assembled (Figure S1c) we perform a final cleaning step using anisole, acetone and IPA, followed by a second annealing in argon, with the same parameters described above. This final annealing step is crucial to remove contaminant and residual PPC, as well as eventual blisters from the heterostructure [2].

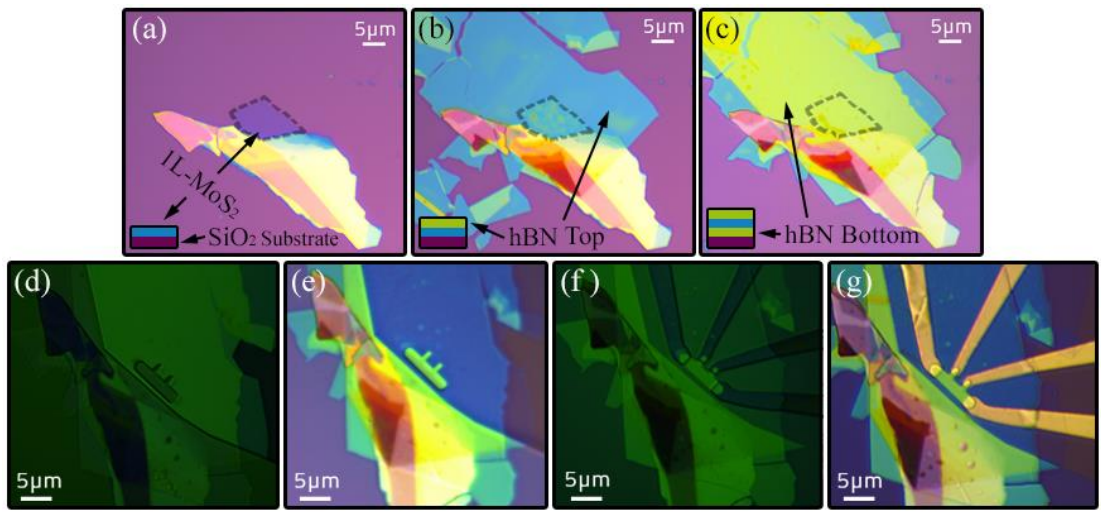

**Supplementary Figure 1.** A typical fabrication process for MoS$_2$ devices: Exfoliation of a SL MoS$_2$ flake (a), transfer of the top layer h-BN on MoS$_2$ flake (b), entire heterostructure (h-BN/MoS2/h-BN) (c). PMMA mask for the geometry definition of the device (d), heterostructure after the dry-etching process (e), Second lithography for contacts and pads (f), final device after the evaporation of titanium/gold and lift off process (g).

Once the h-BN/MoS$_2$/h-BN heterostructure is assembled, the next step is the geometrical definition of the device by electron beam lithography (EBL) with a *Raith Elphy Plus* EBL system. We use a homemade PMMA (4% in chlorobenzene) as resist, spin coated at 4000 rpm for 1 minute and baked at 160 ºC for 10 minutes. The use of chlorobenzene instead of commercial PMMA in anisole permits an easier and very homogeneous resist coating without the need of previous treatment, as well as a thicker coating, useful for the etching mask and the final lift off-process.

After the EBL exposure (electron dose: 250 µC/cm$^2$ at 15 kV), we develop the resist with a mixture of 1 part MIBK to 3 parts of isopropanol, which represents a good compromise between very good contrast and enough sensitivity [3]. The resulting structure is shown in Figure S1d.

Next, (Figure S1e) we etch away the EBL-exposed areas by dry plasma etching with an ICP-RIE Plasma Pro Cobra 100 in SF$_6$ atmosphere (40 sccm, P=75W, process pressure 6 mTorr and T= 10 ºC)[4]. The etching rate is fixed to 2 nm/s and controlled by a DC bias. As further discussed below, the sides of the etched structure have a pyramidal profile, fundamental for the consequent achievement of the edge contacts (Figure S1e). After this etching process, we clean the sample in acetone and IPA and carry a new annealing process, similar to the previous one, to remove residual contaminants of the PMMA resist.

After defining the stack geometry, a second EBL process (electron dose: 270 µC/cm$^2$ at 15 kV) is used to define the contact geometry (Figure S1f). For this step, special care was taken while designing the electrodes to avoid contact with eventual multilayer MoS$_2$ flakes.

Finally, we deposit titanium and gold (5/45 nm) by e-beam evaporation. The evaporation takes place at very low pressure (10$^{-8}$ mbar) in a main chamber with a base pressure of 10$^{-10}$ mbar. The final device, after a lift-off process in acetone, is shown in Figure S1g.

We carry all the fabrication steps for the device in one day. In particular it is very important to avoid any possible oxidation of the edge contacts. For this reason, it is crucial to spend the minimum time between the etching process and the loading into the pre-chamber of the e-beam evaporator.

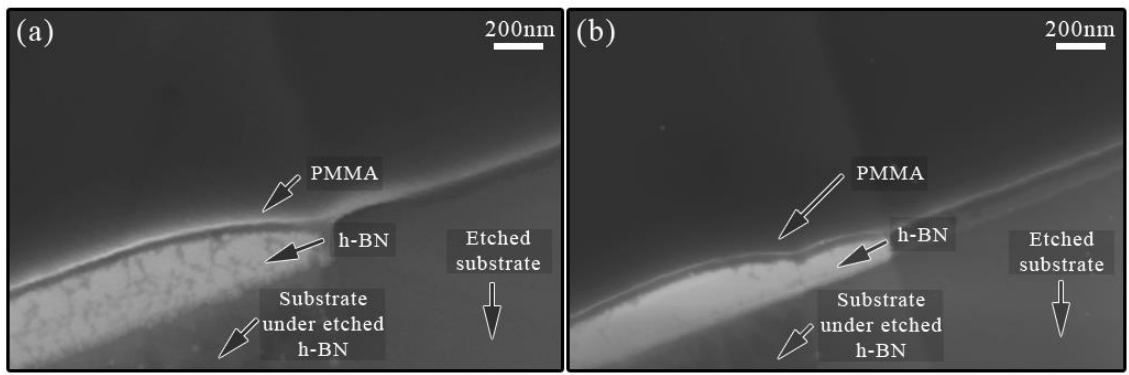

**Supplementary Figure 2.** SEM tilted image of two etching processes. (a) Same etching recipe of Jain et al.[6]. (b) Our etching recipe in SF$_6$ atmosphere.

**Etching process for the edge contacts**

The etching process has a fundamental role for the edge contacts fabrication in high-quality encapsulated devices, where the encapsulation guarantees an isolation from the external environment. In the case of edge contacts, the sides of the etched structure should have a pyramidal profile,[5] as this improves the contact between the atomic layer of $MoS_2$ and the adhesion layer metal (in our case Ti).

In Figure S2 we compare two SEM tilted images for two different etching processes of h-BN. In Figure S2(a) we used the same recipe (in $SF_6$+Ar atmosphere) of Jain et al.[6], in which the authors recently obtained low resistance edge contacts in encapsulated $MoS_2$ devices. In Figure S2(b) we show the etched interface using our etching recipe (only $SF_6$ atmosphere but at very low pressure). In both cases the interface presents a clean pyramidal profile and the substrate shows negligible degradation during the etching process. However, our recipe results in a more homogeneous etching.

**Supplementary Note 2: Raman and photoluminescence characterization**

We determine the thickness of the $MoS_2$ flakes used for device fabrication by a combination of optical microscopy, Raman mapping and Photoluminescence. Figure S3a shows an optical microscope image of the MoS$_2$ flake used to fabricate the device described in the main text, and Figure S3b shows a false color map of the ratio between the summed intensities of the A$_{1g}$ + E$^1_{2g}$ Raman peaks of MoS$_2$ and the intensity of the Si peak, in logarithmic scale. The different thicknesses can be clearly distinguished in the figure. Figure S3c shows individual spectra acquired at the different regions labelled in Figure S3a. The number of layers can be here confirmed by the difference $\Delta f$ between the spectral positions of the E$^1_{2g}$ and A$_{1g}$ peaks [7,8]. For the thinnest region we obtain $\Delta f = 19.4$ cm$^{-1}$, compatible with the values given in literature for 1L-MoS$_2$.

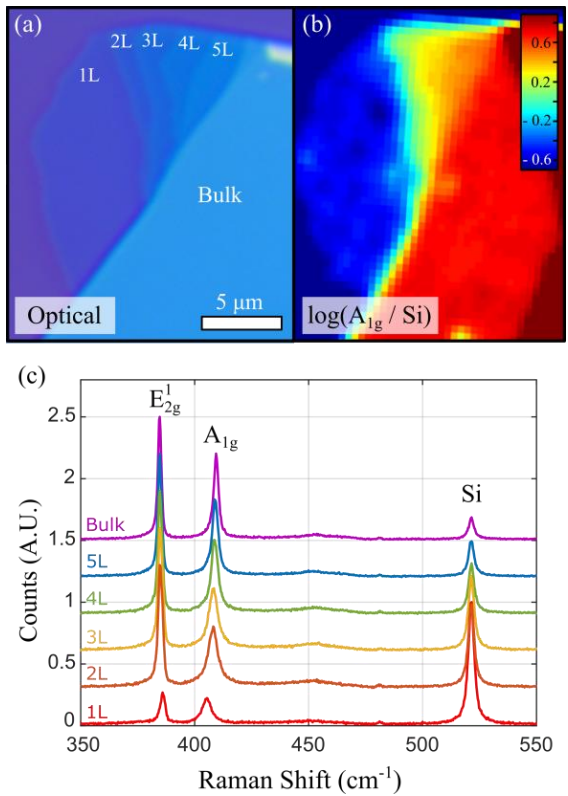

**Supplementary Figure 3.** Raman characterization of the $MoS_2$ thickness. (a) Optical microscopy image of the $MoS_2$ flake used for the device of the main text. The labels indicate regions with different thickness. (b) False color Raman map of the difference between the $A_{1g}$ and Si peak intensities, as labeled in panel c. (c) Raman spectra acquired at the different regions labelled in Figure 1a. The spectra show three prominent peaks corresponding to the $A_{1g}$ and $E^1_{2g}$ Raman modes of $MoS_2$ and the Si Raman mode.

We further confirm the thickness of the MoS$_2$ flakes by measuring the position of the A exciton peak in their photoluminescence spectrum. Figure S4 shows room-temperature photoluminescence spectra acquired at two separate monolayer MoS$_2$ flakes. The $X^A_{1s}$ exciton peak can be clearly observed at around 1.87 eV, in good agreement with the values found in literature [8–10]. For different monolayers we observe small fluctuations (~30 meV) of the A peak position, which we attribute to differences in the relative strength of their exciton and trion transitions.

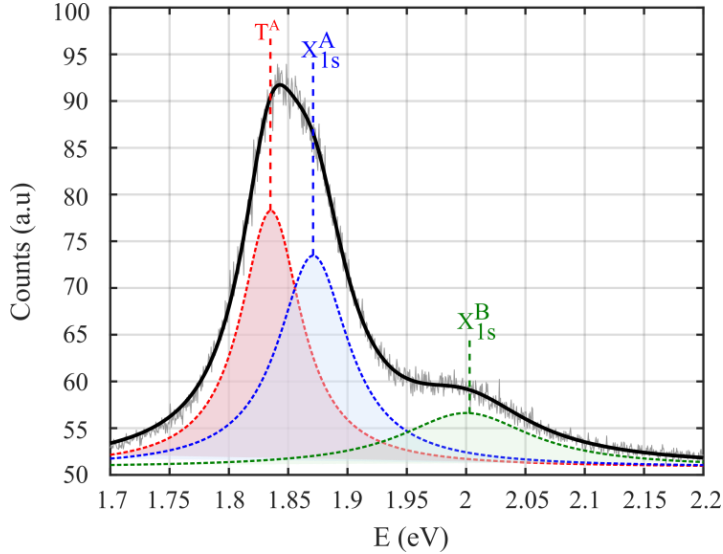

**Supplementary Figure 4.** Room-temperature photoluminescence spectra of monolayer MoS$_2$ on SiO$_2$ under 530 nm excitation. The dashed lines are the individual contributions from the T$^A$, $X^A_{1s}$ and $X^B_{1s}$ exciton transitions.

**Supplementary Note 3: Drain-source voltage dependence of the photocurrent spectra**

Figure S5a shows two *I-V* characteristics measured for the same gate voltage while keeping the 1L-MoS$_2$ device in the dark (orange curve) and under illumination on resonance with the $X^A$ transition. We observe that the effect of exposure to light is similar to increasing the gate voltage, as expected when photoconductivity is dominated by the photovoltaic effect. In particular, for $V_{sd}$ above the saturation current ($V_{sat} \approx 1.7$ V), the photocurrent remains mostly constant, while for the photoconductive effect it should increase as the drain-source electric field becomes larger.

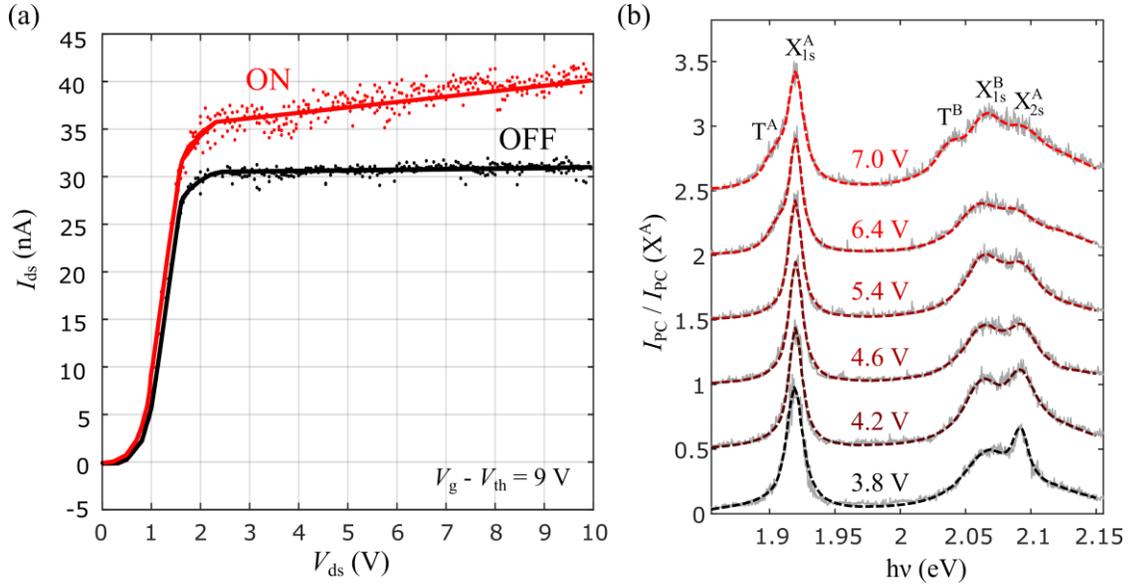

**Supplementary Figure 5.** *I-V* characteristics of the 1L-MoS$_2$ device measured in the dark and under optical excitation on resonance with the X$^A$ transition.

Figure S5b shows PC spectra acquired for at $V_g - V_{th} = 9$ V for different drain-source voltages. We observe that the relative weight of the X$^B$ and T$^B$ features modulate with $V_{sd}$, with T$^B$ becoming larger at higher voltages. A similar trend is observed as well for the X$^A$ and T$^A$ features. Changing the drain-source voltage also allows to modulate the intensity of the X$^A_{2s}$ peak, which becomes larger for low drain-source voltages.

**Supplementary Note 4: Experimental setup for low-temperature photocurrent spectroscopy**

The experimental setup for PCS is schematically depicted in Supplementary Figure 6. The sample is placed inside a pulse-tube cryostat with an optical access at 5K and exposed to laser illumination. The light source is a supercontinuum (white) laser (SuperK Compact), and the excitation wavelength is selected using a monochromator (Oriel MS257 with 1200 lines/mm diffraction grid). This allows to scan the visible and NIR spectral range, roughly from 450 nm to 840 nm. The setup also includes a halogen lamp and a CCD camera, aligned with the laser excitation via two beam splitters, which allows for an easy sample alignment with micrometric resolution. In order to improve the signal-to-noise ratio of the optoelectronic measurements, the excitation signal is modulated by an optical chopper and the electrical response of the device is registered using a lock-in amplifier with the same modulation frequency.

In order to expose the full device to light, we keep the output slit of the monochromator fully open. This results in a broad, rainbow-like, rectangular illumination field with photon energy shifting by 40 meV from side to side of the exposed area (see Supplementary Figure 6b). Because the optically active area of the device, i.e. the 1L-MoS$_2$ area between the drain and source electrodes, is much smaller than the exposed area, it effectively acts as a secondary slit. Thus, the effective bandwidth for the spectral measurements is given by the shift in photon energy from drain electrode to source electrode, $\Delta h\nu \approx 0.2$ meV.

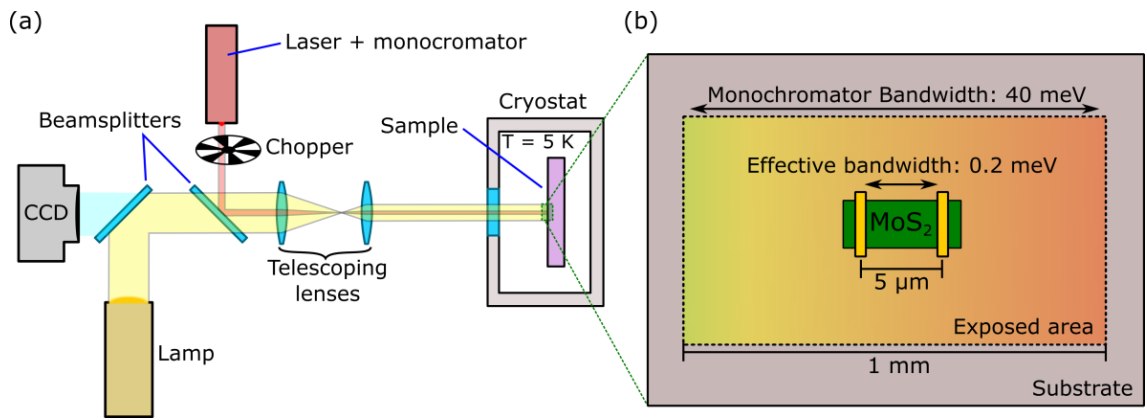

**Supplementary Figure 6.** (a) Experimental setup for photocurrent spectroscopy at cryogenic temperature. (b) Schematic representation of the illumination conditions at the sample. The laser excitation is spread across a 1 mm wide area centred at the position of the device at study, with the excitation energy shifting by 40 meV from side to side of the area.

**Supplementary Note 5: Theoretical modelling and exciton energy**

As discussed in the text, nonlocal screening effects alter the energy levels of 2D hydrogenic excitons, especially the lower ones since the radius is of the order or smaller than the screening length. The electron-hole interaction is found to diverge logarithmically on approaching each other instead of the usual Coulomb behavior of the form $1/r$, where $r = |\mathbf{r}_e - \mathbf{r}_h|$ is the electron-hole separation.

Nonlocal screening effects in thin semiconductor are usually described within the Keldysh approach[37]. The potential that interpolates between the Coulomb potential for large separation and logarithm divergence at small electron-hole distance is given, in CGS units, as follows

$$V(r) = -\frac{\pi e^2}{2r_0} F\left(\frac{\kappa r}{r_0}\right), \qquad (S1)$$

where $\kappa$ is the dielectric constant of the embedding medium, $r_0$ is related to the screening length due to the 2D polarizability of the 1L-MoS$_2$ and

$$F(z) = H_0(z) - Y_0(z). \qquad (S2)$$

Here $H_0$ and $Y_0$ stands for the zeroth order Struve function and Bessel function of the second kind, respectively. The effective-mass equation for the $n$s exciton states with energy $E_n$ is expressed as

$$H\psi_{ns}(r) = (E_n - E_g)\psi_{ns}(r), \tag{S3}$$

in terms of the quasiparticle bandgap $E_g$ and the Hamiltonian $H$ for the relative coordinate $r$

$$H = -\frac{\hbar^2}{2\mu}\frac{d^2}{dr^2} + V(r). \tag{S4}$$

Equation (S3) for the potential (S1) can be diagonalized by standard means. To this end, it is convenient to use dimensionless magnitudes. Thus, we define the dimensionless relative coordinate $z = \kappa r/r_0$. The effective-mass equation then reads

$$\left(-\frac{d^2}{dz^2} - u_0 f(z)\right)\psi_{ns}(z) = \epsilon_n \psi_{ns}(z), \tag{S5}$$

with $u_0 = \pi\mu/m_0\kappa z_B$, and $\epsilon_n = (\mu/m_0 R_y z_B^2)(E_n - E_g)$. Here $z_B = \kappa a_B/r_0$ is the value of the dimensionless coordinate when the electron-hole separation equals the atomic Bohr radius $a_B = \hbar^2/e^2 m_0 = 0.0529$ nm and $R_y = e^4 m_0/2\hbar^2 = 13.605$ eV.

Let $z_{\max}$ be the maximum value of the dimensionless relative coordinate considered in the numerical solution of equation (S5). We now set a grid of $N+1$ equally spaced points, $z_l = lh$, with $h = z_{\max}/N$ and $l = 0,1,\cdots,N$. For brevity we define $f_l = f(lh)$ and $\psi_l^{(n)} = \psi_{ns}(lh)$. Therefore, the discretized effective-mass equation can be cast in the form

$$(2 - h^2 u_0 f_l)\psi_l^{(n)} - \psi_{l+1}^{(n)} - \psi_{l-1}^{(n)} = h^2 \epsilon_n \psi_l^{(n)}, \tag{S6}$$

with boundary conditions $\psi_0^{(n)} = \psi_N^{(n)} = 0$. Hence, the problem is reduced to finding the eigenvalues of a real, symmetric and tridiagonal matrix for a given set of parameters $E_g$, $\kappa$ and $r_0$. Once the lowest eigenvalues $\epsilon_n$ are found, the exciton levels are given by

$$E_n = E_g + \frac{m_0 z_B^2}{\mu}\epsilon_n R_y. \tag{S7}$$

**Supplementary Note 6: Additional spectral measurements**

Figure S6 shows PC spectra for different gate voltages. The peaks of the Rydberg excited states appear in a consistent way for different photocurrent spectra, and their intensities change with the gate voltage in a very non-trivial way. We also observe small fluctuations in the positions of the peak maxima for different gate voltages, which could be caused by changes in the electrostatic screening from the conduction-band electrons and the neighbouring environment. For high gate voltages, the excited states become more prominent compared to the fundamental states.

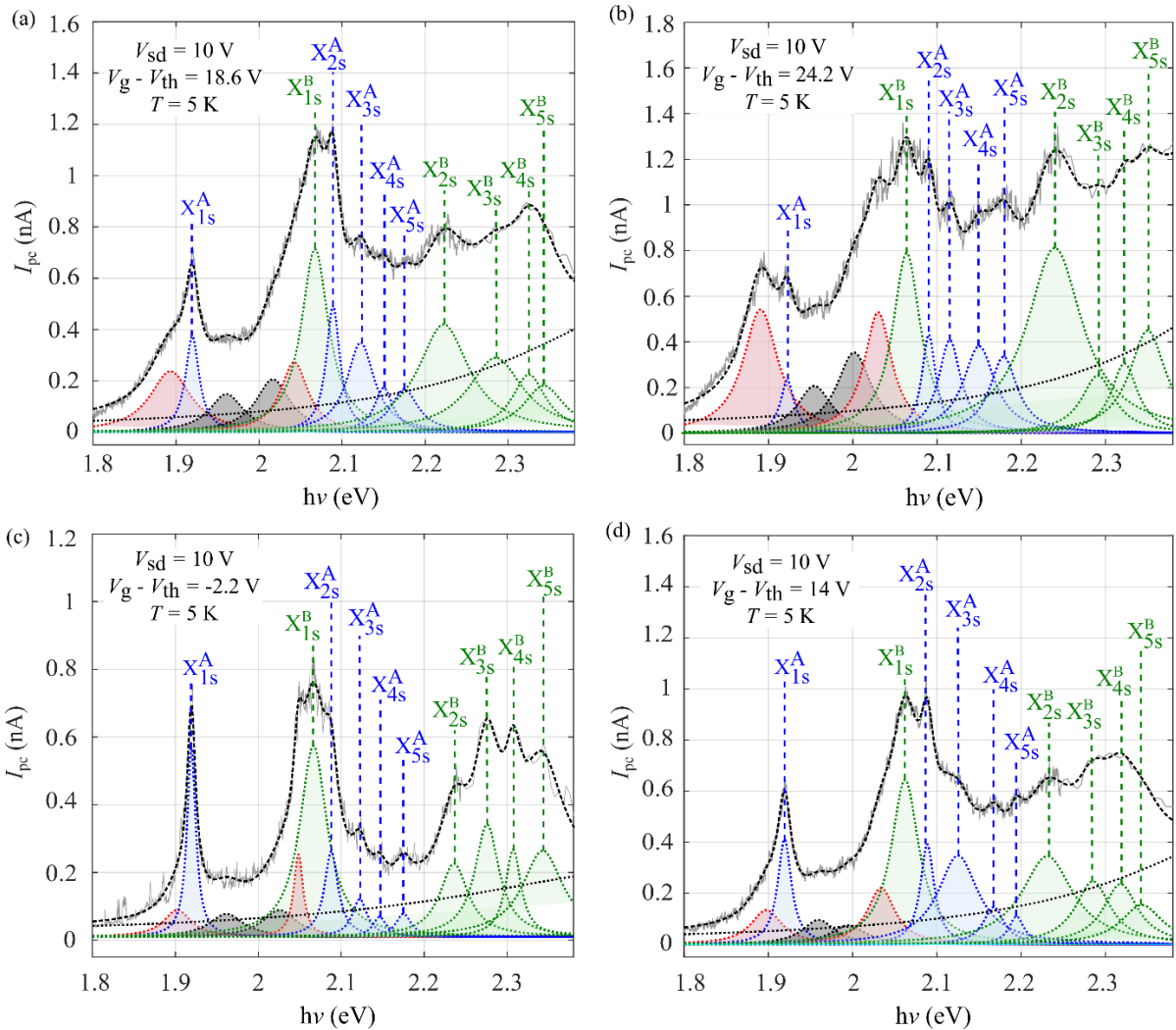

**Supplementary Figure 6**. PC spectroscopy of excited Rydberg states at different gate voltages. Panels (a), (b), (c) and (d) show photocurrent spectra at $V_{sd}$ = 18.6, 24.2, -2.2, 14 V respectively. Experimental data (gray line) are fitted to a multiple-peak Lorentzian plus an exponential background (black, dashed line). The individual Lorentzian functions in blue and green correspond to Rydberg series of excitons A and B respectively.

# Supplementary Note 7: Photocurrent power dependence

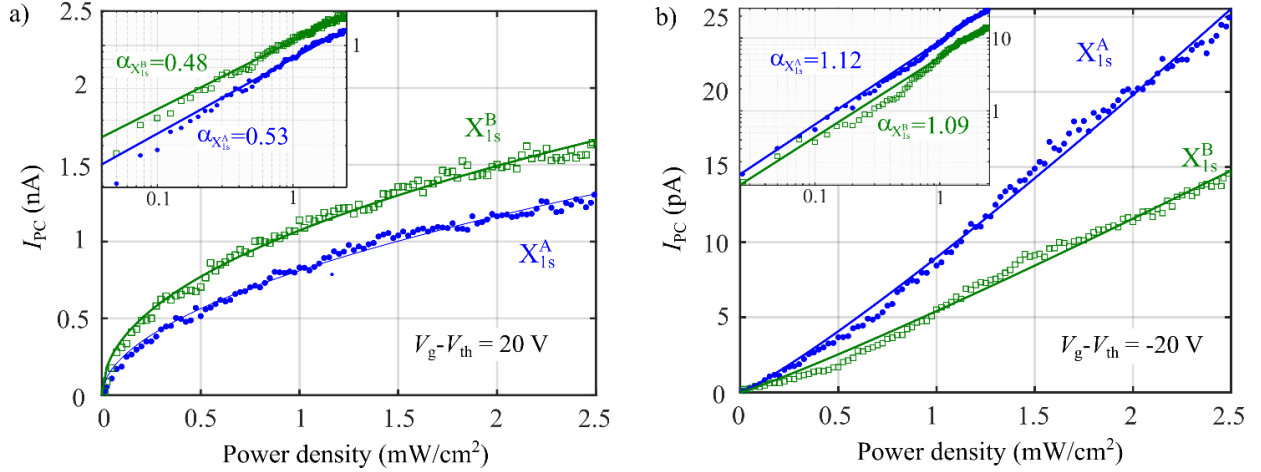

**Supplementary Figure 7**. Power density dependence of the photocurrent for two different gate voltages: $V_g$-$V_{th}$ = 20V (a) and $V_g$-$V_{th}$ = -20V (b), measured for illumination on resonance with $X_{1S}^A$ (blue dots) and $X_{1S}^B$ (green squares). Insets show the same graphs in logarithmic scale.

Supplementary Figure 7a shows the illumination power dependence of the photocurrent for a gate voltage $V_g$-$V_{th}$ = 20V and two different excitation wavelengths, on resonance with the $X_{1S}^A$ and $X_{1S}^B$ transitions. We observe that, for this gate voltage, the photocurrent increases sublinearly with the illumination power $P$, as expected for photovoltaic effects.[11] By fitting the experimental data to $I_{PC} = P^\alpha$ we get $\alpha \approx 0.5$ for both cases. As discussed in the main text, when the gate voltage is set well below $V_{th}$, photovoltaic effects become largely reduced, as they are proportional to the device transconductance (see Figure 3b in the main text). In consequence, the measured photocurrent decreases roughly by a factor 40. Interestingly, when we measure the power dependence for the sub-threshold voltage regime ($V_g$-$V_{th}$ = 20V; Supplementary Figure 7b), we get $\alpha \approx 1$, indicating that the remaining photocurrent is predominantly produced by the photoconductive effect, expected to change linearly with the illumination power.

## Supplementary Note 8: Effect of temperature on the PC spectra

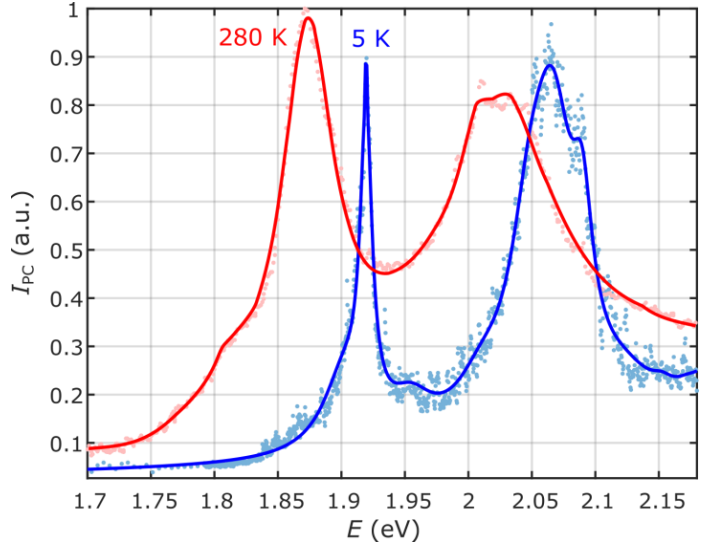

**Supplementary Figure 8**. Photocurrent spectra at T=280 K (red) and T=5K (blue), measured at $V_g$-$V_{th}$= -5V. Solid lines are fits to a multiple-peak Lorentzian functions.

Supplementary Figure 8 shows two photocurrent spectra acquired at $T = 280$ K and at $T = 5$ K for the same $V_g$-$V_{th}$= -5V. We observe two main differences between the spectra: Firstly, for the 280 K spectrum the excitonic spectral features appear red-shifted by 45 meV due to increased electron-phonon interaction[10,12]. Secondly, thermal disorder leads to peak broadening at 280 K. For the $X_{1s}^A$ peak we get a FWHM of 53 meV at $T = 280$ K, roughly 4 times broader than for T = 5 K.
As expected, in the room-temperature PC spectrum presented here the B exciton feature is much more prominent than in PL spectra acquired in similar conditions (see Supplementary Figure 4).

## Supplementary Note 9: Estimation of carrier density and fermi energy shift

In the following we use a capacitor model to estimate the increase in carrier density $\delta n$ produced by the gate voltage. The gate voltage $V_g$, *i.e.* the total voltage drop between the Si back gate and the MoS$_2$ channel, will be given by

$$\delta V_g = \delta E \cdot d + \frac{1}{e}\delta E_F \quad \text{(S8)}$$

Where $E$ is the electric field between the electrode and the flake, $-e$ is the electron charge and $E_F$ is the Fermi energy. For a parallel plate with two different insulator layers the geometrical capacitance is

$$C_g = \left( \frac{d_{SiO_2}}{\epsilon_0 \epsilon_{SiO_2}} + \frac{d_{BN}}{\epsilon_0 \epsilon_{BN}} \right)^{-1}, \quad \text{(S9)}$$

and we have

$$\delta E \cdot d = \frac{e \delta n}{C_g}. \tag{S10}$$

Replacing in (S8) and using $\delta E_F = (\delta E_F/\delta n)\,\delta n = \delta n/D$, where $D$ is the density of states of the 2D semiconductor, we get

$$\delta V_g = \frac{de}{\epsilon_0 \epsilon_d} \cdot \delta n + \frac{1}{eD}\delta n = \left(\frac{1}{C_g} + \frac{1}{C_q}\right) e\delta n, \tag{S11}$$

where we have defined the quantum capacitance as $C_q = e^2 D$. We can now express (S11) in terms of the Fermi energy using $\delta E_F = \delta n/D$. We get

$$\delta V_g = \left(\frac{1}{C_g} + \frac{1}{C_q}\right) eD\, \delta E_F = \left(\frac{1}{C_g} + \frac{1}{C_q}\right)\frac{C_q}{e}\,\delta E_F. \tag{S12}$$

Therefore, solving for $E_F$, we have

$$\delta E_F = \frac{e\delta V_g}{1 + \frac{C_q}{C_g}}. \tag{S13}$$

We model the density of states of 1L-MoSe$_2$ as the step function

$$D(E) = \begin{cases} g_{2D} \equiv \frac{\mu_{\text{eff}}}{\pi \hbar^2} & \text{if } E > E_{\text{CB}} \\ 0 & \text{if } E > E_{\text{CB}} \end{cases}, \tag{S14}$$

where $\mu_{\text{eff}}$ is the electron effective mass in MoS$_2$ ($\mu_{\text{eff}} = 0.35\,m_0$) and $E_{\text{CB}}$ is the edge of the conduction band. Then, by integrating equation (S13) we get

$$\Delta E_F = \frac{e}{1 + \frac{e^2 g_{2D}}{C_g}} (V_g - V_{\text{th}}), \tag{S15}$$

where $V_{\text{th}}$ is the threshold voltage at which $E_F = E_{\text{CB}}$. In our case, we get $\Delta E_F/(V_g - V_{\text{th}}) = 0.28$ meV V$^{-1}$, which for the maximal gate voltages applied here ($V_g - V_{\text{th}} = 50$V) gives $\Delta E_F = 14$ meV. Finally, the density of excess carriers, $n$ can be obtained as $n = \Delta E_F \cdot g_{2D} = 7.17 \times 10^{10}\,\text{cm}^{-2}\text{V}^{-1}(V_g - V_{\text{th}})$. Thus, the maximal carrier densities reached here are of $n = 3.58 \times 10^{12}\,\text{cm}^{-2}$.

**Supplementary Note 10: Gate modulation of the $X^A$-$T^A$ splitting**

The gate-induced modulation of the A exciton-trion splitting in monolayer MoS₂ has been studied in e.g. ref. 17. There, they show that, when a gate voltage $V_g > V_{th}$ is applied to increase the fermi energy $E_F$ of the 1L-MoS₂ channel, the $X^A$ peak position shifts by $E_F$. In our measurements, we only reach fermi energies of, at most, 14 meV (see Supplementary Note 9), much lower than the 50 meV energies reached in ref. 17. Thus, this effect is only observed weakly. Supplementary Figure 9a shows the spectral position of the $X^A_{1s}$ and $T^A$ peaks as a function of $V_g - V_{th}$ and figure 1b shows the exciton-trion energy splitting. For the voltage range between $V_g - V_{th} = 0$ V and $V_g - V_{th} = 36$ V, corresponding to a Fermi energy shift of 10 meV, the exciton-trion splitting increases by roughly 15 meV, compatible with the results from earlier literature.

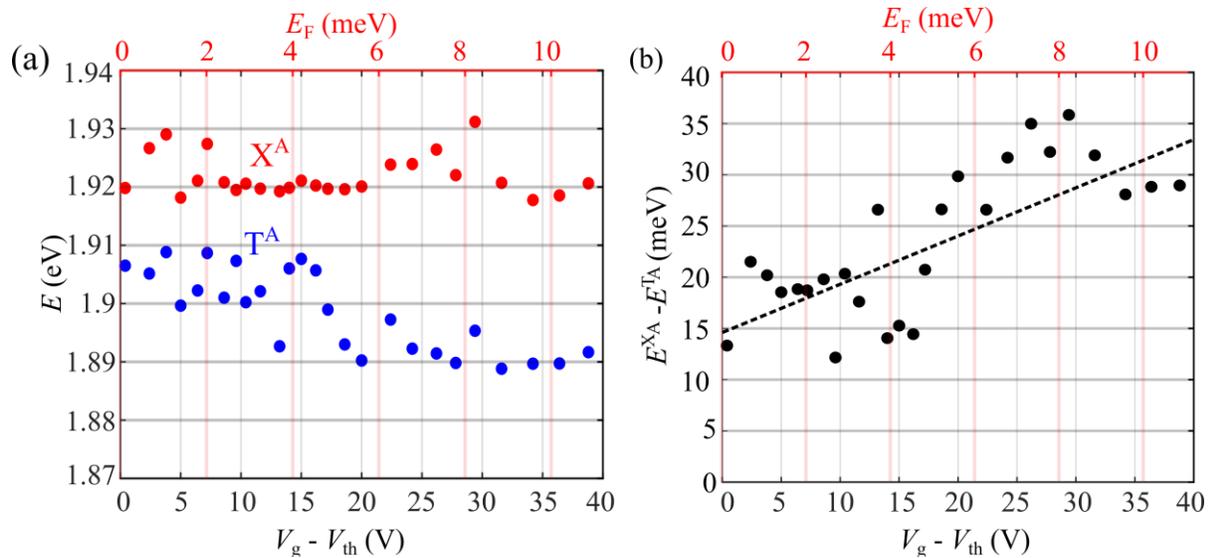

**Supplementary Figure 9**. Gate modulation of the $X^A$-$T^A$ splitting. (a) Spectral position of the $X^A$ and $T^A$ peaks, extracted from multi-lorentzian fittings of the PC spectra measured between $V_g - V_{th} = 0$ V and $V_g - V_{th} = 40$ V. The top axis indicates the corresponding fermi energy, relative to the edge of the conduction band. (b) Energy splitting between the $X^A_{1s}$ and $T^A$ transitions.

**Bibliography**

(1) Pizzocchero, F.; Gammelgaard, L.; Jessen, B. S.; Caridad, J. M.; Bøggild, P.; Wang, L.; Hone, J.; Bøggild, P.; Booth, T. J. The Hot Pick-up Technique for Batch Assembly of van Der Waals Heterostructures. *Nat. Commun.* **2016**, *7*, 11894.

(2) Purdie, D. G.; Pugno, N. M.; Taniguchi, T.; Watanabe, K.; Ferrari, A. C.; Lombardo, A. Cleaning Interfaces in Layered Materials Heterostructures. *Nat. Commun.* **2018**, *9*, 5387.

(3) McCord, M.; Rooks, M. *Handbook of Microlithography, Micromachining, and Microfabrication. Volume 1: Microlithography 1, Chapter2: Electron Beam Lithography, SPIE, 1997.*

(4) Clericò, V.; Delgado-Notario, J. A.; Saiz-Bretín, M.; Malyshev, A. V; Meziani, Y. M.; Hidalgo, P.; Méndez, B.; Amado, M.; Domínguez-Adame, F.; Diez, E. Quantum Nanoconstrictions Fabricated by Cryo-Etching in Encapsulated Graphene. *Sci. Rep.* **2019**, *9*, 13572.

(5) Wang, L.; Meric, I.; Huang, P. Y.; Gao, Q.; Gao, Y.; Tran, H.; Taniguchi, T.; Watanabe, K.; Campos, L. M.; Muller, D. A.; *et al.* One-Dimensional Electrical Contact to a Two-Dimensional Material. *Science (80-. ).* **2013**, *342*, 614–617.

(6) Jain, A.; Szabó, Á.; Parzefall, M.; Bonvin, E.; Taniguchi, T.; Watanabe, K.; Bharadwaj, P.; Luisier, M.; Novotny, L. One-Dimensional Edge Contacts to a Monolayer Semiconductor. *Nano Lett.* **2019**, *19*, 6914–6923.

(7) Li, H.; Zhang, Q.; Yap, C. C. R.; Tay, B. K.; Edwin, T. H. T.; Olivier, A.; Baillargeat, D. From Bulk to Monolayer MoS 2: Evolution of Raman Scattering. *Adv. Funct. Mater.* **2012**, *22*, 1385–1390.

(8) Buscema, M.; Steele, G. a.; van der Zant, H. S. J.; Castellanos-Gomez, A. The Effect of the Substrate on the Raman and Photoluminescence Emission of Single-Layer MoS2. *Nano Res.* **2014**, *7*, 1–50.

(9) Splendiani, A.; Sun, L.; Zhang, Y.; Li, T.; Kim, J.; Chim, C.-Y.; Galli, G.; Wang, F. Emerging Photoluminescence in Monolayer MoS2. *Nano Lett.* **2010**, *10*, 1271–1275.

(10) Christopher, J. W.; Goldberg, B. B.; Swan, A. K. Long Tailed Trions in Monolayer MoS 2: Temperature Dependent Asymmetry and Resulting Red-Shift of Trion Photoluminescence Spectra. *Sci. Rep.* **2017**, *7*, 1–8.

(11) Furchi, M. M.; Polyushkin, D. K.; Pospischil, A.; Mueller, T. Mechanisms of Photoconductivity in Atomically Thin MoS2. *Nano Lett.* **2014**, *14*, 22.

(12) Tongay, S.; Zhou, J.; Ataca, C.; Lo, K.; Matthews, T. S.; Li, J.; Grossman, J. C.; Wu, J. Thermally Driven Crossover from Indirect toward Direct Bandgap in 2D Semiconductors: MoSe2 versus MoS2. *Nano Lett.* **2012**, *12*, 5576–5580.